%% file: config-granular.tex
\documentclass[acmtog,authorversion,nonacm,screen]{acmart}

\AtBeginDocument{%
  }

\input{preamble}

\begin{document}

\title{The Configurational Element Method for Nonconvex Granular Media}


\author{Zhecheng Wang}
\email{zhecheng@cs.toronto.edu}
\orcid{0000-0003-4989-6971}
\affiliation{%
  \institution{University of Toronto}
  \city{Toronto}
  \country{Canada}
}

\author{Breannan Smith}
\email{breannan@cs.toronto.edu}
\orcid{0009-0000-2789-2123}
\affiliation{%
  \institution{University of Toronto}
  \city{Toronto}
  \country{Canada}
}

\author{Abhishek Madan}
\email{amadan@cs.toronto.edu}
\orcid{0000-0002-6119-7500}
\affiliation{%
  \institution{University of Toronto}
  \city{Toronto}
  \country{Canada}
}

\author{Eitan Grinspun}
\email{eitan@cs.toronto.edu}
\orcid{0000-0003-4460-7747}
\affiliation{
    \institution{University of Toronto}
    \city{Toronto}
    \country{Canada}
}

\begin{abstract}
Granular media surround us, comprising everything from the ground we walk on to the foods we eat. Owing to their ubiquity our ability to understand and predict the mechanical evolution of grains is not only of key scientific importance, but is also a key component to synthesize believable animations of our world. Despite their importance, shortcomings persist in our ability to simulate granular media. In particular, simulating grains with non-convex shapes remains a challenging and computationally expensive task. We propose a method to simulate non-convex rigid grains by posing geometric contact in configuration space and learning the resulting contact map with a neural network. Our formulation reduces the complex task of modeling and simulating non-convex shapes to simple network evaluations that are easily run on standard compute hardware, allowing us to quickly and robustly simulate large scale systems of non-convex grains.
\end{abstract}


\begin{teaserfigure}
    \centering
    \includegraphics[width=\linewidth]{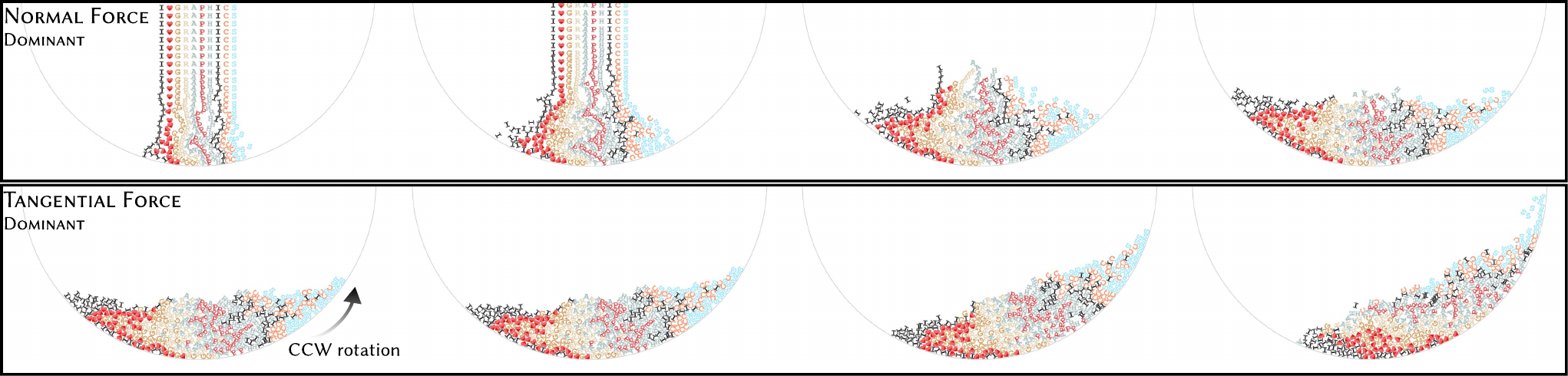}
    \caption{\textbf{Spinning Drum}. We fill a drum with grains that spell I heart G R A P H I C S. As the drum spins, we induce contact between each pair of grain shapes.}
    \label{fig:spinning_drum}
\end{teaserfigure}

\maketitle

\input{sections/introduction.tex}
\input{sections/related_works.tex}
\input{sections/theory.tex}
\input{sections/neural_distance_field.tex}
\input{sections/implementation.tex}
\input{sections/results.tex}
\input{sections/discussion.tex}
\bibliographystyle{ACM-Reference-Format}
\bibliography{references}

\input{sections/appendix/appendix.tex}
\input{sections/appendix/multi_contact_theory.tex}

\end{document}

%% file: preamble.tex
\PassOptionsToPackage{table}{xcolor}
\usepackage{bm}
\usepackage{amsmath}
\usepackage{booktabs}
\usepackage[abs]{overpic}
\usepackage{nicefrac}
\usepackage{wrapfig}
\usepackage{soul}
\usepackage{colortbl}
\usepackage{duckuments}
\usepackage[ruled,vlined]{algorithm2e}

\definecolor{turquoise}{cmyk}{0.65,0,0.1,0.1}
\definecolor{purple}{rgb}{0.65,0,0.65}
\definecolor{darkgreen}{rgb}{0.0, 0.5, 0.0}
\definecolor{darkred}{rgb}{0.5, 0.0, 0.0}
\definecolor{darkblue}{rgb}{0.0, 0.0, 0.5}
\definecolor{blue}{rgb}{0.0, 0.0, 1.0}
\definecolor{orange}{rgb}{1.0, 0.5, 0.0}
\definecolor{red}{rgb}{1.0, 0.0, 0.0}
\definecolor{cherry}{RGB}{186,12,47}
\definecolor{pink}{RGB}{117,107,177}
\definecolor{oursgray}{gray}{0.90}

\newcommand{\vect}[1]{\mathbf{#1}}

\newcommand{\refsec}[1] {Section~\ref{#1}}
\newcommand{\refeq}[1] {Equation~\ref{#1}}
\newcommand{\reffig}[1] {Fig.~\ref{#1}}

\newcommand{\reftab}[1] {Table~\ref{#1}}
\newcommand{\refapx}[1] {Appendix~\ref{#1}}

\newcommand{\dx}{\nabla_{\vect{x}_B}}
\newcommand{\dtheta}{\nabla_{\theta_B}}
\newcommand{\dxA}{\nabla_{\vect{x}_A}}

\newcommand{\SEtwo}{\mathrm{SE}(2)}

%% file: sections/introduction.tex
\section{Introduction}

Granular media are one of the most common arrangements of solid matter; in industrial applications, they are the second-most commonly handled material behind water~\cite{Richard:2005:Slow}. In daily life, examples include ensembles of rocks, food grains, snow, sand, and coal. While the pairwise contact interactions between grains are simple, these interactions quickly produce complex emergent phenomena as the number of grains increases. Understanding these effects is essential for studying natural phenomena such as landslides, designing structures such as crop silos and pharmaceutical powders, and simulating large collections of objects for visual effects, games, and robotics.

Grain-level simulations are challenging, and most prior work makes significant simplifying assumptions, most commonly modelling every grain as a disc or a sphere~\cite{cundall1979discrete,Poschel:2005}. While this greatly simplifies contact handling, it limits applicability in cases where grains are non-convex or highly anisotropic, and lifting these restrictions typically makes simulations prohibitively slow~\cite{kawamoto2016levelset,lim2014contact,rakotonirina2019grains3d}.

A recurring strategy in rigid body simulation and computer graphics has been to \emph{precompute} expensive geometric interactions and cache them for fast runtime queries~\cite{barbicjames2007_timecritical,barbicjames2008_6dofhaptics,jamespai2004_bdtree,zhang2014continuous}. This idea has appeared in many forms, including precomputed contact response, cached impulse responses, and reduced models for collision and deformation, enabling simulations to trade memory for speed. Such approaches are particularly effective when the space of possible interactions is far smaller than the number of interaction events encountered during simulation.

This observation is especially relevant for granular media. Even when grains are geometrically complex, it is common for the number of distinct grain shapes to be far smaller than the number of grains themselves. For example, a small variety of manufactured parts may be poured into a large bin, or natural grains classifiable into a few discrete shape categories~\cite{Powers1953} may fill a silo or beach. Under this premise, given \(n\) grains and only \(m\) grain varieties with \(m \ll n\), there are only ${m \choose 2} + m$ distinct geometric interaction pairs, even though the number of contacts that occur over time is orders of magnitude larger.

This separation exposes a natural opportunity for precomputation: the geometric complexity of the simulation is entirely contained in a small set of interaction pairs, while the runtime simulation repeatedly queries those interactions at many different relative configurations. The central challenge is therefore not whether to precompute, but \emph{how}: each interaction pair admits a continuum of relative configurations (in our case, all relative poses in \(\SEtwo\)), and naively storing contact information for all such states is infeasible. Moreover, as the simulation increasingly relies on precomputed data, memory bandwidth becomes a dominant cost; compact representations are essential if precomputation is to reside in fast memory and remain effective at scale. Representing contact interactions directly in configuration space provides one such avenue for achieving this compactness. 

In this work, we show that for rigid grains in two dimensions, all contact and frictional behaviour between a pair of shapes can be reduced to just two scalar fields defined over relative configuration space.

Early foundational work showed that contact forces and frictional responses can, in principle, be expressed and reasoned about geometrically in configuration space, including for multiple simultaneous contacts~\cite{erdmann1993multiple,wriggers2006contact}. However, in this general form, configuration-space contact formulations involve high-dimensional contact descriptions and combinatorial contact modes, making them difficult to precompute and store compactly across all relative configurations in large-scale, many-body simulation settings.

Our approach relies on two modelling choices whose role is to collapse the high-dimensional, combinatorial structure of contact into a compact, fixed-dimensional representation suitable for precomputation. The first choice is standard in penalty-based rigid body simulation: we posit that the configuration-space distance between two objects admits a well-defined gradient in a neighbourhood of the contact manifold. Although the exact object--object distance is not differentiable along medial axes, any simulation based on penalty forces or barrier energies effectively mollifies the distance field~\cite{ferguson2021intersection,lan2022affine}. Under such smoothing, gradients are well-defined almost everywhere and provide stable directions for contact forces and torques.

The second choice concerns frictional response and is more restrictive. We posit that, for any interacting pair in a given relative configuration, the projection of the contact moment arm onto the contact normal direction is uniquely defined. This holds trivially for single-point contacts, and we show that it also holds for common extended-contact regimes such as flat face-on-face contact. While this assumption does not hold universally for arbitrary non-convex geometries, it allows frictional forces and torques to be represented without explicitly tracking multiple contact points, enabling a compact configuration-space representation.

Under these two assumptions, all geometric information required for contact and friction between a given pair of grain types can be encoded by just two scalar-valued functions over relative configuration space: a signed distance function and a normal-projected moment-arm function. We represent these functions using neural fields, enabling the complete precomputation for a single interaction pair, over all relative configurations in \(\SEtwo\), to be stored in just tens of kilobytes.

Together, these design choices transform contact handling from an online geometric problem into a data-driven lookup problem with fixed cost per interaction. This shift enables granular simulations with heterogeneous, non-convex grains to scale to tens of thousands of bodies while retaining rich, shape-dependent contact and frictional behaviour, without explicit geometric contact computation at runtime.

We demonstrate that this approach enables large-scale simulations of non-convex granular media in two dimensions, producing stable and physically plausible results across a range of challenging scenarios.

\paragraph{Contributions.}
This paper presents a configuration-space approach to simulating granular media with heterogeneous, non-convex grains, designed for regimes in which the number of grain varieties is far smaller than the number of grains. By precomputing and compactly encoding pairwise geometric interactions, the method enables efficient large-scale simulation without explicit geometric contact queries. Our contributions are as follows:

\begin{itemize}
  \item \textbf{A configuration space contact formulation for non-convex grains.}
  We formulate grain--grain contact using a signed distance function defined over relative rigid-body configurations, allowing contact detection and normal-force computation to be decoupled from explicit contact point tracking, even in the presence of multi-point and extended contact.

  \item \textbf{A compact friction model based on normal-projected moment arms.}
  We introduce a configuration-space friction formulation that represents frictional forces and torques using a single contact normal and a scalar normal-projected moment arm. This model captures frictional behaviour for a range of practically important contact regimes while avoiding explicit multi-contact bookkeeping.

  \item \textbf{Neural representations of precomputed contact interactions.}
  For each pair of grain geometries, we learn two neural fields over relative configuration space: a signed distance field and a projected moment-arm field. Together, these fields encode all geometric information required for contact and friction in just a few kilobytes per contact type.

  \item \textbf{An efficient SDF-based method to generate configuration space training data.}
  We present a practical procedure for approximating object--object distances, penetration depths, and virtual contact points using only object-space signed distance fields, enabling scalable generation of training data concentrated near the contact manifold.

  \item \textbf{A granular media simulation built on learned configuration space interactions.}
  We integrate the proposed contact and friction models into a discrete element simulation that combines a standard broad phase with learned narrow-phase queries, supporting heterogeneous, non-convex grain types without explicit geometric contact resolution.

  \item \textbf{Demonstrations of non-convex granular behaviour.}
  We demonstrate the physical behaviour of the configuration-space friction model on canonical rigid-body tests and more complex two-dimensional simulations involving over ten thousand non-convex grains, exhibiting shape-dependent effects such as interlocking, jamming, and changes in angle of repose.
\end{itemize}

%% file: sections/related_works.tex
\section{Related Work}
Our work primarily builds upon granular media simulation and rigid body simulation in both computational mechanics and computer graphics, and learning-based methods for these simulation methods.

\subsection{Granular Media Simulation}
Granular media involves modelling a large ensemble of particles as individual rigid bodies that interact with each other.
Such methods originated from the molecular dynamics literature~\cite{Alder:1957,Alder:1959,Alder:1960}, and were later extended to granular media in geomechanics~\cite{cundall1979discrete}, rheology~\cite{Walton:1986}, and other physics disciplines~\cite{Haff:1986,Gallas:1992}.
Later work also explored grain fracturing~\cite{Nguyen:2015:Numerical,Aastrom:1998:Fragmentation}, and some more recent work has explored hybrid particle-continuum approaches that judiciously use particles only where needed~\cite{yue2018hybrid,chen2021hybrid}.
The two dominant forces in such simulations, contact and friction, can be modelled as penalty forces~\cite{cundall1979discrete}, smoothed particle hydrodynamics (SPH) forces~\cite{Ivain:2011}, or hard constraints via a mixed linear complementarity (MLCP) problem~\cite{moreau1994granular}.
The simplicity of the former comes at a price of time-consuming parameter tuning to obtain stable and physically accurate results that respect geometric interpenetration constraints without excessive force magnitudes; meanwhile, the mathematical guarantees of the latter make MLCP methods significantly less efficient~\cite{Kaufman:2008}.
As such, penalty methods are often used in practice, and with sufficient tuning are trustworthy enough to be used to validate continuum models of the same materials~\cite{Rycroft:2009}.
Our simulation is based on the penalty-based discrete element method.

\subsection{Non-Spherical and Non-Convex Grain Contact}
A major shortcoming of most prior work in granular media simulation is that the grains are modelled with spherical geometry.
This significantly simplifies the contact model but is not realistic for all materials and particles, particularly those that are highly anisotropic and/or non-convex such as medicine pills, legumes, and organic molecules.
Spherical geometry has only one contact state (up to rotational symmetry), but convex non-spherical geometry can have several contacting configurations, and non-convex geometry can even have multiple contact points or surfaces in one configuration.
In this most general case, computing contact points is an ongoing challenge, and most commonly used methods fail to produce high-quality contact points in all cases~\cite{erleben2018methodology}, with commonly used software libraries such as Bullet providing convex proxies for complex geometry~\cite{Coumans2021}.
Some work that explores non-convex grains, e.g., via NURBs surfaces~\cite{lim2014contact}, unions of convex shapes~\cite{rakotonirina2019grains3d} or grid-based level sets~\cite{kawamoto2016levelset}, is limited to small simulations of only thousands of grains; in our work we demonstrate simulations over 11{,}000 non-convex grains. 
Recent work derives homogenized continuum models of non-convex grains from small, grain level simulations~\cite{chen2025numericalhomogenization}, but does not capture large scale grain level interactions, is limited to isotropic grain shapes, and does not treat non-uniform heterogeneous mixtures of grain shapes.

\subsection{Configuration-Space Contact and Planning}
A substantial body of work in robotics and graphics has studied contact reasoning directly in configuration space.
Early foundational work showed how contact forces and frictional responses can be expressed and reasoned about geometrically in configuration space, including for multiple simultaneous contacts~\cite{erdmann1993multiple}.
Subsequent work explored contact tracking and mode transitions in configuration space for haptic rendering and compliant motion~\cite{rosell2005haptic}.
Other approaches precompute or approximate the boundary of the contact space to accelerate proximity queries such as penetration depth, including configuration-space distance metrics~\cite{zhang2007cdist,zhang2008fast}, sampled contact-space propagation~\cite{he2016efficient}, and continuous parameterizations of the contact manifold~\cite{zhang2014continuous}.
These methods demonstrate the power of configuration-space formulations, but typically focus on geometric proximity queries rather than reusable contact response models suitable for large-scale multi-body simulation.

Related to these configuration-space formulations, a separate line of work has studied motion planning in regimes dominated by tight contact.
For example, Zhang et al.~\cite{zhang2020cspace} propose a tunnel-discovery strategy for rigid-body disentanglement puzzles, explicitly reasoning about narrow passages in configuration space induced by contact constraints.
While this work focuses on planning rather than contact force or frictional response, it highlights the importance of configuration-space structure in scenarios where motion is tightly constrained by contact, and refers to a broader range of sampling-based and geometric planning approaches developed for such settings.

\subsection{Learning-Based Approaches}
Machine learning approaches have consistently proven to produce compact and effective representations of complex functions, making them well-suited for abstracting away the complex and expensive geometric queries at the heart of rigid body and granular media simulation.
Furthermore, such representations are often smooth and easily differentiable, which adds useful regularization properties on top of the geometric data that they model which is often merely continuous, and simplifies crucial operations such as obtaining contact normals.
These approaches involve either learning the geometry for each object~\cite{park2019deepsdf} for accelerating graph-based simulations~\cite{rubanova2024learning}, or learning a more abstract configuration space parameterizing the positions of two objects undergoing contact.

Early work used support vector machines to learn penetration depth or contact boundaries in configuration space~\cite{pan2013efficient,zhang2014continuous}.
More recent approaches increasingly rely on neural networks, particularly in robotics, for learning swept-volume or collision-related fields in configuration space~\cite{joho2024neural,guo2025deepcollide,li2024configuration}, though a ``pose'' in this context typically refers to joint-angle coordinates rather than rigid body configurations.
Other work learns local collision fields directly in workspace, such as triangle--triangle collision predictors parameterized by vertex positions~\cite{zesch2023neural}.

Most closely related to our approach, Cai et al.~\cite{cai2022active} learn a neural signed distance field in configuration space to represent self-collision boundaries for reduced deformable models.
Their method demonstrates that neural implicit representations can effectively encode contact manifolds in configuration space, but is limited to self-collision, reduced deformation spaces, and does not model frictional response or inter-object contact.
Some recent granular media approaches have also applied machine learning to collision detection~\cite{huang2024machine,lai2022machine}, though these methods are restricted to convex geometries.
To our knowledge, our work is the first to use neural networks to learn reusable contact information, including frictional response, between arbitrary pairs of non-convex grains directly in rigid body configuration space.

%% file: sections/theory.tex
\section{Rigid-Body Kinematics}
\label{sec:kinematics}
For Sections~\ref{sec:kinematics}, \ref{sec:contact_model}, and~\ref{sec:friction_model}, we consider two rigid bodies $A$ and $B$, with corresponding states $q_A$ and $q_B$ in \(\SEtwo\).
The corresponding group action of a state $q$ on a vector $\mathbf{p} \in \mathbb{R}^2$ (i.e., the rigid body transformation encoded in $q$ applied to $\mathbf{p}$) is denoted by $q(\mathbf{p})$.
In two dimensions, we represent such states as a transformation from a body frame to a world frame by a rotation by angle $\theta$ followed by a translation of the center of mass $\mathbf{x} = [x,y]^\top$; concatenated together a rigid body state is represented as a tuple $q = (\theta, \mathbf{x})$ or equivalently a vector in $\mathbb{R}^3$, $\mathbf{q} = [\theta, x, y]^\top$ (the vector representation may also be written as $[\theta, \mathbf{x}^\top]^\top$).
We subscript these quantities to denote the object they reference.
Under this encoding, we write the group action as
\begin{equation}
    \label{eq:rb_2d}
    q(\mathbf{p}) = R(\theta)\mathbf{p} + \mathbf{x},
\end{equation}
where $R(\theta) = \left[ \begin{array}{c c} \cos \theta & -\sin\theta \\ \sin\theta & \cos\theta \end{array} \right]$.

We also define \textit{relative} states that, for example, transform a body point in $A$ to a body point in $B$ (and similarly from $B$ to $A$).
The group structure of $\SEtwo$ implies that each state $q$ has an inverse $q^{-1}$, so the relative state from $A$ to $B$ is $q_{AB} = q_A^{-1}q_B$ (concatenation is the group operation applied right to left).

The states and their components vary with time, and therefore $q_A$ and $q_B$ have associated world-space body velocities $\dot{q}_A$ and $\dot{q}_B$, respectively, which encode each body's rotational velocity and linear (translational) velocity.
We write the velocity for a state as $\dot{q} = (\omega, \mathbf{v})$ or $\dot{\mathbf{q}} = [\omega, \mathbf{v}^\top]^\top$ in vector form (subscripted as appropriate).
We are also interested in the velocities at specific points on the surface, which combine linear and angular velocity into a single vector: for a world-space point on $A$ denoted by $\mathbf{p}_A$, the point's velocity is $\mathbf{v}_{\mathbf{p}_A} = \mathbf{v}_A + \omega_A R(\pi/2)(\mathbf{p}_A - \mathbf{x}_A)$.
When $A$ and $B$ touch at a point, associated with $A$ and $B$ via $\mathbf{p}_A$ and $\mathbf{p}_B$ respectively, the relative velocity at the contact point of $B$ with respect to $A$ is $\mathbf{u} = \mathbf{v}_{\mathbf{p}_B} - \mathbf{v}_{\mathbf{p}_A}$.
The same notion of relative point velocities applies unchanged between pairs of closest points on $A$ and $B$.


\section{Configuration-Space Contact Model}
\label{sec:contact_model}
\begin{figure}
    \centering
    \includegraphics[width=\columnwidth]{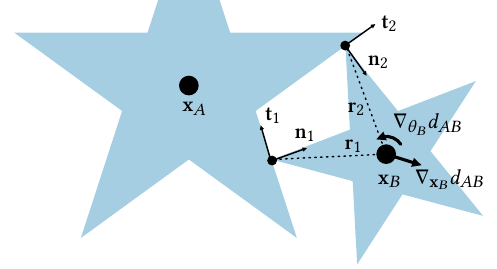}
    \caption{A two-point contact between two stars, and the resulting translation and rotation gradients. The gradients are drawn as a vector and rotation, respectively, on the center of mass of $B$.}
    \label{fig:contact-model}
\end{figure}

The key ingredient to a contact model is a signed distance function that measures the separation between two objects and defines an admissible region for their configurations where the distance is non-negative.
Contact occurs at the zero level set of this function, and the gradients of this distance function are used to compute forces and torques on each object.
With a linear approximation of the contact surface (i.e., a contact plane), the resulting forces act along each contact normal.
Unlike prior approaches that explicitly track multiple contacts and produce contact normals at each point, we abstract away the individual contact points and our configuration-space formulation yields a single contact normal for \textit{all} contact points between $A$ and $B$.
At configurations with multiple contact points between objects, the distance field exhibits gradient discontinuities (an object-object analogue to medial axes that occur in point-object signed distance functions), but if the field is smoothed (e.g., via convolution with an arbitrarily small isotropic Gaussian), the gradient discontinuities disappear, while smooth regions are unaffected.
Therefore, we make the assumption that the distance field always yields a well-defined gradient.
The advantage of this formulation is that we can abstract away the full geometry at each configuration and simply use the object configurations in $\SEtwo$ as input to the distance function for mediating contact between objects.
In this section, we outline a configuration-space distance function and a geometric approach to reason about the contact normals induced by such a function.
We show a didactic multi-contact example in \reffig{fig:contact-model} and we discuss the implementation in \refsec{sec:implementation}.

\subsection{Distance Formulation}

We aim to describe a minimum distance function $d_{AB}(q_A,q_B)$ between the two rigid bodies, as a function of their states.
Notably, the distance function encodes the \textit{geometry} of $A$ and $B$, hence the subscript, which enables this decoupling between geometry and rigid body state.
Further note that this distance is \textit{signed}, where negative distances encode the penetration depth.
The (non-negative) distance is the solution to the following optimization problem:
\begin{equation}
    \label{eq:dist_opt}
    d_{AB}(q_A, q_B) = \min_{\mathbf{p}_A \in A, \mathbf{p}_B \in B} \| q_A(\mathbf{p}_A) - q_B(\mathbf{p}_B) \|,
\end{equation}
while the (signed) penetration depth is the (negative) length of the smallest vector that separates two objects under translation:
\begin{align}
    \label{eq:pd_opt}
    -\textrm{PD}_{AB}(q_A, q_B) =& \min_{\mathbf{t} \in \mathbb{R}^2} \| \mathbf{t} \| \\
    & \text{s.t. } d_{AB}(q_A, (\theta_B, \mathbf{x}_B + \mathbf{t})) \ge 0. \nonumber
\end{align}
The distance is continuous with respect to $q_A$ and $q_B$.

\subsection{Contact Normals}

Ultimately, we want to use $d_{AB}$ to not only detect near-intersecting and intersecting configurations, but to also resolve collisions.
For the latter task, we require the gradient of $d_{AB}$, particularly near the zero level set of $d_{AB}$.
In the following, we discuss the gradient for body $B$, but the same approach applies for $A$.
Translation gradients, denoted $\dx d_{AB}$, provide a direction for normal forces; and rotation gradients, denoted $\dtheta d_{AB}$, provide a direction and moment arm-scaling for normal torques.
Under the smoothing assumption, it is sufficient to describe the behaviour of the gradient in a single-contact case, though the geometric interpretation of the gradient under multi-point contact remains the same (see \refapx{app:multi_contact}).
The translation gradient $\dx d_{AB}$ is simply the (Euclidean) contact normal at the contact point, which is the unit-length vector that produces the maximal increase in distance.
Similarly, the rotation gradient $\dtheta d_{AB}$ is the rate at which the contact point separates from its position, under either a counter-clockwise or clockwise rotation (positive or negative sign, respectively), which is the component of the moment arm orthogonal to the contact normal.
More concretely, for a contact point $\mathbf{p}$, contact normal $\mathbf{n} = \dx d_{AB}$, tangent basis vector $\mathbf{t} = R(\pi/2)\mathbf{n}$, the rotation gradient is $\dtheta d_{AB} = -(\mathbf{p} - \mathbf{x}_B) \cdot \mathbf{t}$.

Combined, the state gradient $\nabla_\mathbf{q} d_{AB} = [\dtheta d_{AB}, (\dx d_{AB})^\top]^\top$ can be used to simultaneously compute the force and torque induced by contact.
However, note that these gradients do not provide magnitudes of the the normal force and torque in configuration space on their own, but merely vectors that can be scaled by the force magnitude; the scale factor will be described in \refsec{sec:implementation}.

\section{Configuration-Space Friction Model}
\label{sec:friction_model}

Along with normal forces, contact also produces friction forces along the contact plane (i.e., tangent to each surface).
In the basic Coulomb friction model (e.g., ~\cite{Moreau:1983}), when two objects are not separating at the contact point, they mutually exert equal and opposite forces along the contact tangent plane (line in $\mathbb{R}^2$).
For a contact point $\mathbf{p}$ with contact normal $\mathbf{n} = \dx d_{AB}$, the friction force is $\mathbf{f}_{f} = \alpha \mathbf{t}$, where $\mathbf{t} = R(\pi/2)\mathbf{n}$ is a unit vector orthogonal to the contact normal that acts as a basis for the tangent line, and $\alpha$ is the force magnitude and signed if the force goes opposite to $\mathbf{t}$.
Similarly, the torque induced by friction at each $\mathbf{p}$ is $\tau_{f} = \mathbf{r} \times \mathbf{f}_{f} = (R(\pi/2)\mathbf{r}) \cdot (\alpha \mathbf{t}) = \alpha(\mathbf{r} \cdot \mathbf{n})$.
These forces and torques are summed over all contact points.

\begin{wrapfigure}{r}{0.2\textwidth}
\centering
\includegraphics[width=0.2\textwidth]{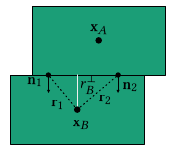}
\end{wrapfigure}
In configuration space, the absence of per-contact tangent planes, and individual moment arms at each contact point, makes it impossible to produce friction forces in the same manner.
The gradient assumption provides us with a well-defined tangential projection of one contact's moment arm, but not the normal components of all contacts as required by friction torques.
To efficiently enable this computation, we make a further assumption that \textit{all} moment arms have the same projection onto the normal.
This is not universally true, but it generalizes many multi-point contact and continuous contact surface scenarios, such as when one box rests on top of another (see inset).
Under this assumption, the remaining geometric ingredient for supporting friction a \textit{normal-projected moment arm function}, defined as $r_B^\perp = \mathbf{r} \cdot \mathbf{n}$ for any moment arm $\mathbf{r}$ in the configuration with respect to $B$.

\begin{figure}
    \centering
    \includegraphics[width=\linewidth]{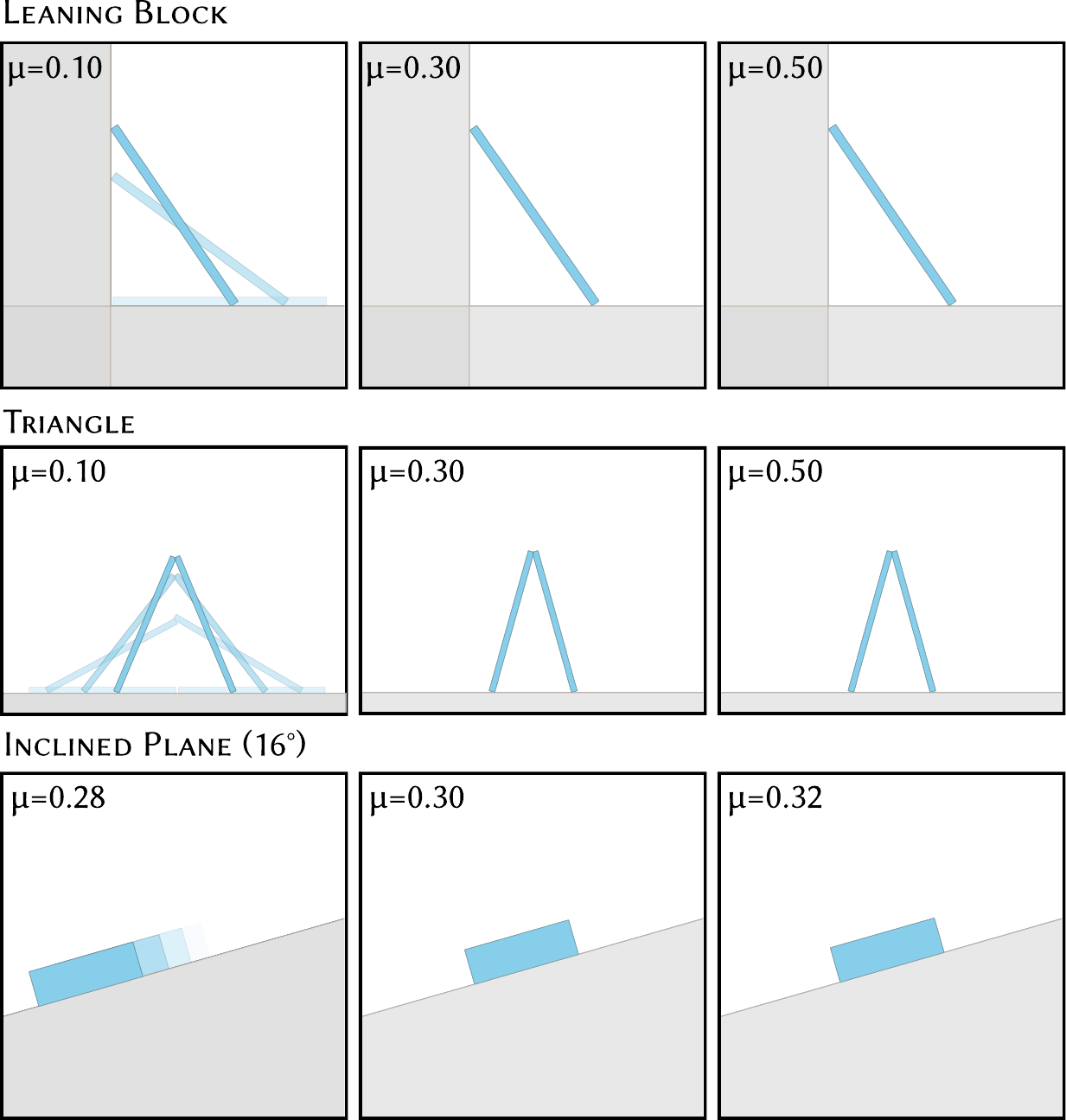}
    \caption{\textbf{C-Space Friction Model Validations.} (1) A block leaning on a corner (2 contact points between 2 objects);
    (2) Two blocks leaning toward each other like a triangle (3 contact points between 3 objects when also counting the ground plane);
    (3) Block on an inclined plane (continuous regions of contact).}
    \label{fig:friction_validation}
\end{figure}

%% file: sections/neural_distance_field.tex
\section{Neural Distance Fields}
\label{sec:neural_dist_field}

The algorithmic complexity of the configuration-space distance functions (\refsec{sec:contact_model}) and projected moment arm functions (\refsec{sec:friction_model}) described earlier, combined with the required smoothing, makes the exact computation a bottleneck in narrow phase computation.
Instead, we learn neural fields as differentiable approximations to these functions that can be compactly stored and quickly evaluated at simulation time.
We train one distance cache $\hat{d}_{AB}$ and one projected moment arm cache $\hat{r}^\perp_{AB}$ per grain geometry pair $(A,B)$; for $m$ types of grains this results in $2\cdot ({m\choose 2} + m)$ networks.
We isotropically rescale all shapes to fit within a unit sphere in $\mathbb{R}^2$ and encode all input angles in the range $[-\pi, \pi]$.
We model all functions with a small multilayer perceptron (MLP) network using rectified linear unit (\texttt{ReLU}) activations, and we train all networks in PyTorch.
We list the architectural details for each experiment (\refsec{sec:results}) in \reftab{tab:experiment_setup}.

\subsection{Minimum Distance/Penetration Depth Cache}
\label{sec:dist_net}
We reduce the input dimension of both the distance networks $\hat{d}_{AB}$ and the projected moment arm networks $\hat{r}_{AB}^\perp$ from $\mathbb{R}^6$ to $\mathbb{R}^3$ by transforming all state to the relative configuration $q_{AB}$, which considers $B$'s state relative to $A$.
To transform the gradients $\nabla_{\mathbf{q}_{AB}} \hat{d}_{AB}$ to this coordinate system, we use the Jacobians $J_B = \frac{\partial \mathbf{q}_{AB}}{\partial \mathbf{q}_B}$ and $J_A = \frac{\partial \mathbf{q}_{AB}}{\partial \mathbf{q}_A}$.
We fit the distance caches $\hat{d}_{\psi;AB}$ separately by learning network weights $\psi$ using a least-squares loss over a dataset of configurations with precomputed signed distances:
\begin{equation}
\label{eq:loss_dist}
\mathcal{L}_d(\psi) = \mathbb{E}_{\mathbf{q} \in \mathcal{D}}\Big[\big(\hat d_{\psi;AB}(\mathbf{q})-d_{AB}(\mathbf{q})\big)^2\Big].
\end{equation}

We discuss the construction of the dataset $\mathcal{D}$ in \refsec{sec:csample}.

\subsection{Projected Moment Arm Cache}
\label{sec:proj_moment_arm_net}
For friction and torque models (\refsec{sec:friction_model}), we also require the normal-projected moment-arm magnitudes $\tilde{\mathbf{r}} \cdot \mathbf{n}$. Since $d_{AB}$ does not provide contact points explicitly, we define a single virtual \textit{normalized} normal from the learned distance field $\hat d_\theta$, $\hat{\mathbf{n}}(\mathbf{q})=\frac{\nabla_{\mathbf{x}}\hat d_\theta(\mathbf{q})}{\|\nabla_{\mathbf{x}}\hat d_\theta(\mathbf{q})\|}$, and use it only to form training targets. (The smoothing of the network means that some regions have non-unit-length translation gradients.) Let $\tilde{\mathbf{r}}_{A},\tilde{\mathbf{r}}_{B}$ denote the (world-space) moment arms from each center of mass to its associated (virtual) closest/contact point; we train a second network $\hat r^\perp_{\phi;AB}$ parameterized by $\phi$ to predict the scalar projections of both $A$ and $B$, $[-\mathbf{r}_{A}\!\cdot\!\hat{\mathbf{n}},\ \mathbf{r}_{B}\!\cdot\!\hat{\mathbf{n}}]^\top$, using a least-squares loss:
\begin{equation}
\label{eq:loss_proj_r}
\mathcal{L}_r(\phi) = \mathbb{E}_{\mathbf{q} \in \mathcal{D}}\left[ \left\| \hat r^\perp_{\phi;AB}(\mathbf{q})-\left[-\mathbf{r}_{A}\!\cdot\!\hat{\mathbf{n}}(\mathbf{q}),\ \mathbf{r}_{B}\!\cdot\!\hat{\mathbf{n}}(\mathbf{q})\right]^\top \right\|_2^2 \right].
\end{equation}

\subsection{Fast Configuration-Space Contact Point Sampling}
\label{sec:csample}

In order to effectively train the networks in \refsec{sec:dist_net} and \refsec{sec:proj_moment_arm_net} for collision detection and response, they must both produce accurate results near the zero-level set of $d_{AB}$, and also reproduce a virtual contact point that leads to plausible forces and torques.
Furthermore, the simulation algorithm has a broad phase that will filter out most object pairs except when they are near this zero level set (\refsec{sec:broad_phase}), so we only need accurate reconstruction in a narrow band near the zero level set.
Therefore, we construct the dataset distribution $\mathcal{D}$ so that 90\% of its samples are near the zero level set of $d_{AB}$ and the remaining 10\% of samples are uniformly distributed in the contact radius that cannot be filtered by the broad phase (\refsec{sec:broad_phase}).
To obtain near-surface samples, we use rejection sampling within a distance radius of 0.1 of the zero level set.
Computing exact object-object distances and penetration depths for non-convex objects can be a significant bottleneck in training, so we introduce a fast approximation that recovers both the minimum distance (or penetration depth) and a closest-point pair using only object-space signed distance fields (SDFs).

\paragraph{Approximation construction.}
Let $\varphi_{A}(\mathbf{x})$ and $\varphi_{q(B)}(\mathbf{x})$ denote the signed distance fields of objects $A$ and $B$ after $B$ is transformed by $q$.
We define the sampling objective
\begin{equation}
\label{eq:sdf_sampler}
\mathcal{E}(\mathbf{x}) = \underbrace{\varphi_A(\mathbf{x}) + \varphi_{q(B)}(\mathbf{x})}_{\text{distance / penetration}}
\;+\;
\underbrace{\big|\varphi_A(\mathbf{x}) - \varphi_{q(B)}(\mathbf{x})\big|}_{\text{absolute difference}},
\end{equation}
and select
\begin{equation}
\mathbf{x}^\star = \arg\min_{\mathbf{x} \in \mathbb{R}^2} \mathcal{E}(\mathbf{x}).
\end{equation}
The two SDFs are overlaid on a $201 \times 201$ grid, and the minimum is selected among the sampled grid points.

\paragraph{Rationale.}
The sum term $\varphi_A+\varphi_{q(B)}$ recovers the minimum separation between the two objects; when the objects interpenetrate, this quantity becomes negative and directly measures penetration depth.
However, minimizing $\varphi_A+\varphi_{q(B)}$ alone is generally \emph{non-unique}: all points on the medial axis of the summed field can achieve the same minimum value, yielding an ill-posed contact location.

The absolute difference term $\lvert\varphi_A-\varphi_{q(B)}\rvert$ acts as a regularizer that biases the solution toward points where the two SDFs are locally balanced, i.e., where the distances to the two surfaces are as equal as possible.
This eliminates the pairs along the medial axis that do not correspond to closest points, and produces a stable SDF value pair $(\varphi_A,\varphi_{q(B)})$ corresponding to an effective closest-point/contact configuration.

\paragraph{Recovered quantities.}
The minimizer $\mathbf{x}^\star$ (the grid point with the minimum value of $\mathcal{E}$) provides:
(i) the minimum distance or penetration depth via $\varphi_A(\mathbf{x}^\star)+\varphi_{q(B)}(\mathbf{x}^\star)$, and
(ii) an implicit closest-point association from which moment arms and projected quantities used in \refsec{sec:friction_model} can be consistently computed.
For configurations that still have multiple minima, we return an arbitrary closest point pair.

It is worth noting that SDF-based approximations to penetration depth lose accuracy for deeply penetrating configurations where the penetration depth can be greater than the global minimum of the SDF, but such states cannot be reached in practice due to contact forces pushing objects apart upon contact.

%% file: sections/implementation.tex
\begin{table*}[ht]
    \centering
    \rowcolors{2}{white}{gray!20} 
    \begin{tabular}{c|c|c|c|c|c|c}
    \textbf{Examples} & \textbf{Grain Count} & \textbf{Grain Types} & \textbf{B.C.} & \textbf{Time Step} $\Delta t$ & \textbf{Fric. Coeff.} $\mu$ & \textbf{NN. Params.}\\
    \textbf{Leaning Block} ($\triangleright$ \reffig{fig:friction_validation}) & $3 (2)$ & $2 (1)$ & Dirichlet & $0.001$ & $0.10/0.30/0.50$ & N/A \\
    \textbf{Triangle} ($\triangleright$ \reffig{fig:friction_validation}) & $3 (1)$ & $2 (1)$ & Dirichlet & $0.001$ & $0.10/0.30/0.50$ & N/A \\
    \textbf{Inclined Plane} ($\triangleright$ \reffig{fig:friction_validation}) & $2 (1)$ & $2 (1)$ & Dirichlet & $0.001$ & $0.28/0.30/0.32$ & N/A \\
    \textbf{Column Collapse} ($\triangleright$ \reffig{fig:column_collapse}) & $6990 (690)$ & $1$ & Dirichlet & $0.00005$ & $0.4$ & $5 \times 64$ \\
    \textbf{Silo Discharge} ($\triangleright$ \reffig{fig:silo_discharge}) & $11311 (1411)$ & $1$ & Dirichlet & $0.00005$ & $0.4$ & $5 \times 64$ \\
    \textbf{Column Packing} ($\triangleright$ \reffig{fig:column_packing}) & $1152 (152)$ & $2(1)$ & Dirichlet & $0.001$ & $0.1$ & $5 \times 64$\\
    \textbf{Spinning Drum} ($\triangleright$ \reffig{fig:spinning_drum}) & $1574 (547)$ & $10(1)$ & Moving & $0.001$ & $0.1$ &$3 \times 32$ \\
    \textbf{Cereal Pouring} ($\triangleright$ \reffig{fig:cereal_pouring}) & $385(105)$ & $5(1)$ & Moving & $0.001$ & $0.7$ & $5 \times 64$\\
    \textbf{Coiling Chain} ($\triangleright$ \reffig{fig:chain_drop}) & $600$ $(101)$ & $1$ & Moving & $0.000025$ & $0.4$ & $5 \times 64$ \\

    \end{tabular}
    \caption{We report, for all examples, the total grain count with the fixed grain count in parentheses, the total number of grain shape types with the number of boundary types (halfplanes and static/kinematic grains) in parentheses, the boundary condition type, the simulation time step, the coefficient of friction, and the number of layers and the neuron count in the contact map MLP. Note that N/A indicates the experiments were conducted without a neural network. Fixed grains participate in collision detection and exert forces on moving grains, typically for the purpose of enforcing boundary conditions, but their state is not updated during time integration.}
    \label{tab:experiment_setup}
\end{table*}

\section{Configurational Space Simulation}
\label{sec:implementation}
We implement a rigid body simulation algorithm that realizes the configuration-space contact model derived in the preceding sections.
The previous sections only describe the ingredients for narrow phase queries between nearby objects, but large-scale simulations also require a broad phase pass to filter out the majority of object pairs that do not interact.
We will first fully describe the narrow phase queries, then zoom out and describe the broad phase pass.

\subsection{Narrow Phase Collision Response}
\label{sec:narrow_phase}
We treat contact and friction with the model proposed by \citet{cundall1979discrete} (also see \citet{chen2021hybrid}), which is the standard model in the granular media literature.
We will briefly describe the model; pseudocode is given in \refapx{app:alg}.

We model normal contact as a penalty force, where the magnitude is proportional to the amount of interpenetration between colliding bodies $A$ and $B$.
This force is given as $\mathbf{f}_{n,B} = k_n d \mathbf{g}_B$ where $d$ is the penetration depth magnitude, $k_n$ is the penalty coefficient, and $\mathbf{g}_B = J_B \nabla_\mathbf{q} \hat{d}_{AB}$.
We model friction with a spring and dashpot formulation, where the spring is placed at the initial point of contact and moves as the motion transitions between slip and stick modes.
The friction force under this model is $\mathbf{f}_{f,B} = -k_t s \mathbf{t}$, where $s$ is the length of the spring, $k_t$ is the penalty coefficient, and $\mathbf{t}$ is a 90-degree counterclockwise rotation of the translation component of $\mathbf{g}_B$.
We increment the spring length $s$ at each timestep by integrating the relative velocity of the contact point in the tangent direction.
The \citet{cundall1979discrete} model also includes viscous damping, which adds a force in the normal and tangent directions that opposes the relative velocity at the contact point, making the final forces $\tilde{\mathbf{f}}_{n,B} = \mathbf{f}_{n,B} - \gamma_n \mathbf{u}^\perp$ and $\tilde{\mathbf{f}}_{f,B} = \mathbf{f}_{f,B} - \gamma_t \mathbf{u}^\parallel$ using damping coefficients $\gamma_n$ and $\gamma_t$.
Despite the lack of an explicit moment arm, the projected moment arm cache gives enough information to compute the tangential component of the relative velocity: $u^\parallel = \mathbf{v}_B - \mathbf{v}_A + \hat{r}^\perp_{AB} \cdot [\omega_A, \omega_B]^\top$.
We compute the relative velocity in the normal direction directly from distance gradients: $u^\perp = \mathbf{g}_B \cdot \dot{\mathbf{q}}_B + \mathbf{g}_A \cdot \dot{\mathbf{q}}_A$.

Beyond the aforementioned parameters for the force computations, we also need a time step $\Delta t$ and a coefficient of friction $\mu$ that determines the width of the friction cone.
We report all simulation and network parameters in \reftab{tab:experiment_setup}.

\subsection{Broad Phase Collision Detection}
\label{sec:broad_phase}
We enclose each rigid body in a bounding sphere centered at its center of mass, with a radius equal to the maximum distance from the center of mass to the object geometry.
We perform broad-phase collision culling with pairwise sphere overlap tests.
For a pair of bodies $A$ and $B$ with states $q_A, q_B \in \SEtwo{}$ and bounding sphere radii $R_A$ and $R_B$, interaction is only possible when the distance between their centers of mass satisfies
$\| \mathbf{t}_A - \mathbf{t}_B \| \le R_A + R_B,$
where $\mathbf{t}_A$ and $\mathbf{t}_B$ denote the translational components of $q_A$ and $q_B$.
Outside of this region, the configuration-space distance $d_{AB}(q_A,q_B)$ is strictly positive and does not depend on the object geometry.

Accordingly, we restrict both the training and the evaluation of the neural configuration-space distance field (\refsec{sec:neural_dist_field}) to the reduced domain
$(\theta, \mathbf{t}) \in S^1 \times \{ \mathbf{t} \in \mathbb{R}^2 \mid \| \mathbf{t} \| \le R_A + R_B \},$
where $\theta$ denotes the relative orientation and $\mathbf{t} = \mathbf{t}_B - \mathbf{t}_A$ the relative translation.

%% file: sections/results.tex
\section{Results}
\label{sec:results}

\subsection{Configuration-Space Friction Model Validation}
\label{sec:friction_validation}

We validate our configuration-space friction model (\refsec{sec:friction_model}) using three canonical rigid-body test cases (\reffig{fig:friction_validation}) that probe distinct frictional regimes. All experiments in this section use the \textit{numerical} configuration-space contact and friction formulation, rather than its neural approximation, to isolate the physical behavior of the model without approximation error.

\paragraph{Leaning Block.}
A single rigid block is placed in contact with two orthogonal half-planes, forming a corner constraint with two active contact points. The block is initialized at a rotation angle of $\theta = 0.6$ radians ($\approx 34.4^\circ$), which is significantly larger than the critical angle corresponding to moderate friction coefficients. We test $\mu \in \{0.1, 0.3, 0.5\}$. For $\mu = 0.1$, the block slides immediately. For $\mu = 0.3$ and $\mu = 0.5$, the block also slides but at substantially reduced rates, with $\mu = 0.3$ exhibiting noticeably slower motion than $\mu = 0.1$. This behavior is consistent with the fact that $\theta$ exceeds the critical angle $\theta_c \approx \arctan(0.3) \approx 16.7^\circ$, while still demonstrating friction-dependent dissipation under multi-point contact.

\paragraph{Triangle.}
Two identical blocks are symmetrically placed leaning toward each other on a ground plane, forming a triangular configuration with three simultaneous contact points: two block--ground contacts and one block--block contact. Each block is inclined at $15^\circ$. Under this configuration, only $\mu = 0.1$ leads to collapse, while larger friction coefficients result in a stable structure. Despite the small inclination angle, the presence of multiple coupled contacts allows frictional forces to redistribute through the contact network, stabilizing the system. This case validates friction resolution under coupled multi-body contact constraints.

\paragraph{Inclined Plane.}
A single block is placed on an inclined plane at $16^\circ$, forming a continuous region of contact. For a block on an incline, sliding occurs when $\tan(\alpha) > \mu$. At $\alpha = 16^\circ$, the critical friction coefficient is $\mu_c \approx \tan(16^\circ) \approx 0.287$. Consistent with this relation, the block remains stationary for $\mu = 0.3$ and $\mu = 0.32$, while sliding initiates at $\mu = 0.28$. This experiment confirms that the model accurately reproduces classical stick--slip behavior in the \textit{continuous} contact regime.

\subsection{Column Collapse}
\begin{figure*}[ht]
    \centering
    \includegraphics[width=\linewidth]{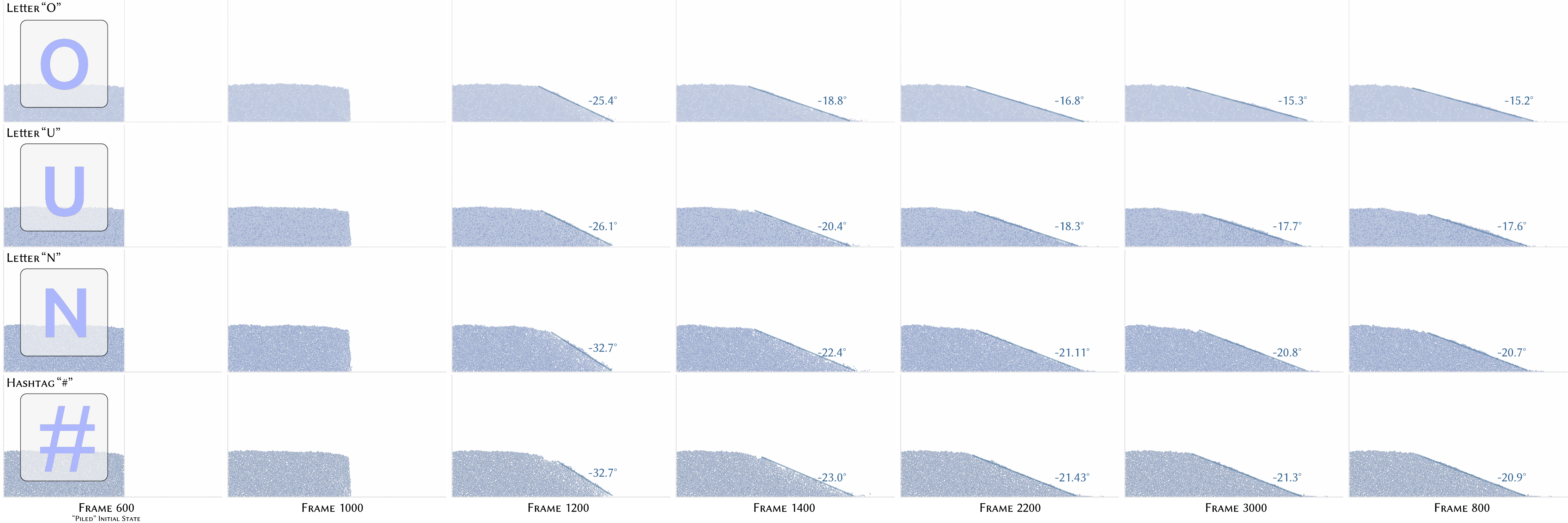}
    \caption{\textbf{Collapsing columns} for four grain types in increasing order of grain complexity and non-convexity. Notably, during and after the collapse, the angle of the front of the collapse increases in tandem with the degree of complexity.}
    \label{fig:column_collapse}
\end{figure*}

We begin our exploration of the influence of grain shape and convexity with a granular column collapse experiment. We fill four rectangular containers with four grain shapes (O, U, N, and \#), and after the transient motion dampens out, we remove a wall to induce an avalanche of grains. See \reffig{fig:column_collapse}. We observe two notable shape dependent effects. First, the smooth shapes, O and U, demonstrate a higher density of grains in the collapsing front compared to the non-smooth grains, N and \#. Second, we observe a difference in the profile shapes: as we increase the degree of non-convexity and shape complexity from O, to U, to N, to \#, we observe an attendant increase in the angle of repose. We simulated all column collapses with the same coefficient of friction $\mu=0.4$.

\subsection{Silo Discharge}
\begin{figure*}[ht]
    \centering
    \includegraphics[width=\linewidth]{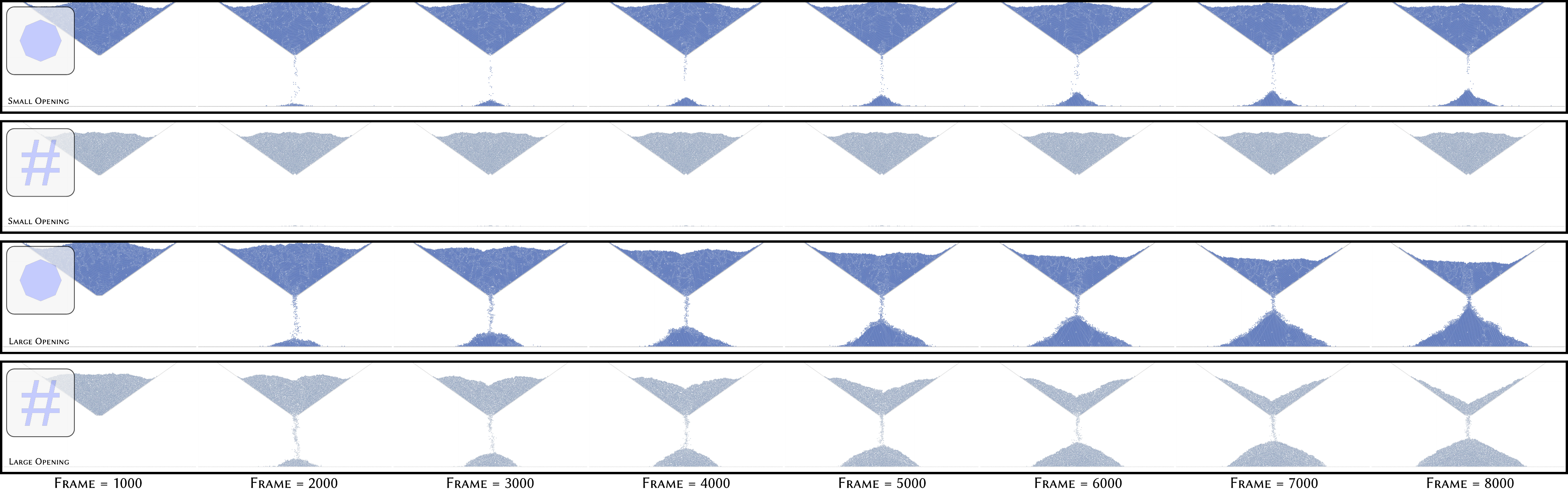}
    \caption{\textbf{Silos discharging grains} for two grain shapes, an octagon and a \#, and a small and large opening width, $8m$ and $16m$. For the small opening with octagons, we observe a slow but steady discharge of grains. In contrast, with \# grains and a small opening, the flow immediately jams. For the large opening, the silo discharges grains for both grain types, but the resulting piles have markedly different profiles. See~\reffig{fig:profile_compare}. Note that the density difference between the octagon and \# shape produces an apparent color shade difference.}
    \label{fig:silo_discharge}
\end{figure*}

\begin{figure}[h]
    \centering
    \includegraphics[width=\linewidth]{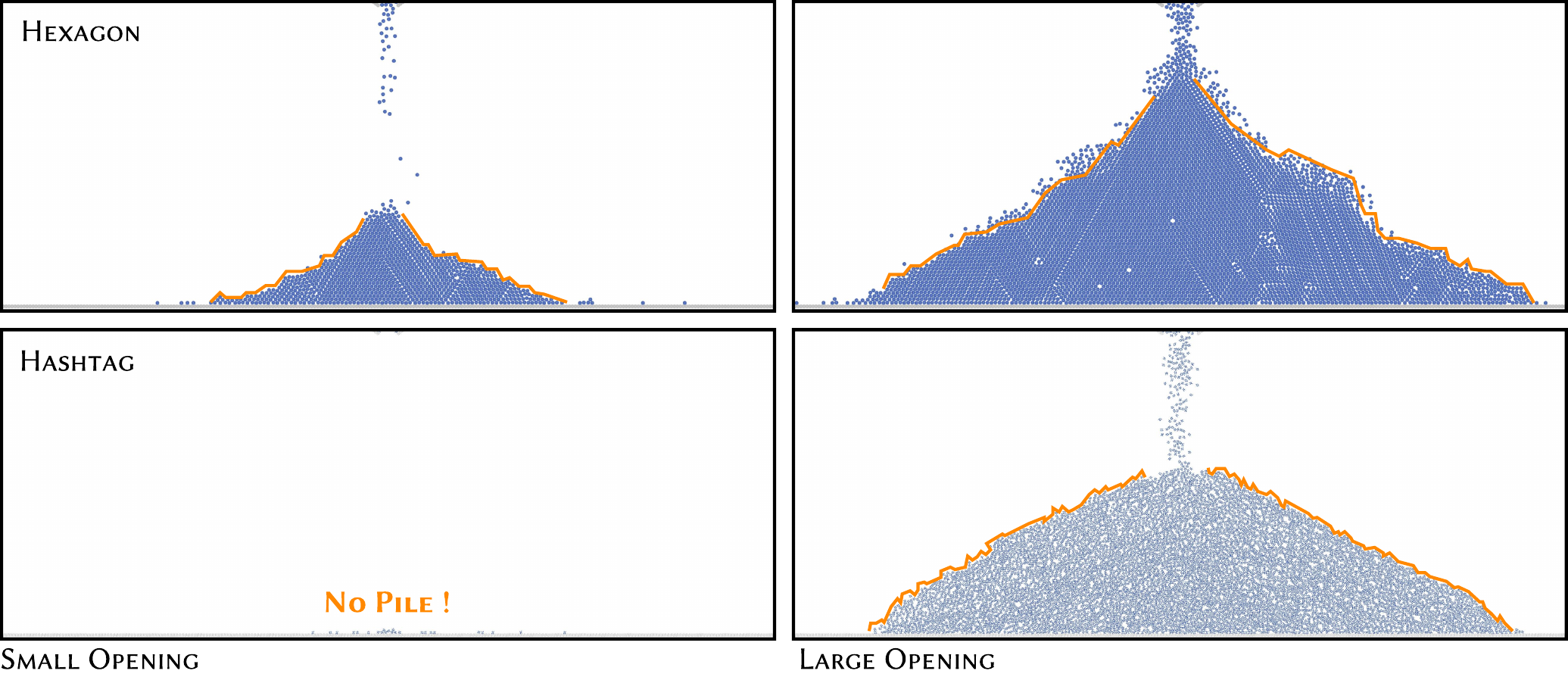}
    \caption{\textbf{Piles of grains formed by a silo discharge}. For an octagon shaped grain type, we observe a low frequency, concave up profile with a sharp peak and a secondary plateau. For the \# shaped grain type, we observe a more gradual concave down dome-like profile with high frequency surface oscillations.}
    \label{fig:profile_compare}
\end{figure}

To further explore the influence of grain shape and convexity on the macroscopic evolution of granular assemblies, we fill silos of multiple funnel widths with convex, octagonal shaped grains and non-convex, hashtag shaped grains, see \reffig{fig:silo_discharge}. Once full, we remove the plugs from the silos' funnels and allow the grains to drain. With a small funnel width and octagonal grains, the silo drains slowly but steadily. With a small funnel width and hashtag shaped grains, in contrast, the funnel immediately clogs and there is no flow. Differences in model parameters do not lead to the jamming of the silo, as we simulate both systems with identical contact parameters; instead, the hashtag's non-convex protrusions interlock, allowing them to form a stable arch more readily than the convex octagons. With a larger funnel opening, both grain types freely drain from the silo, but the resulting piles have different profile shapes. These results show that the grain shape itself plays a strong role in the evolution of granular assemblies.

\subsection{Column Packing}
\begin{figure}[ht!]
    \centering
    \includegraphics[width=\linewidth]{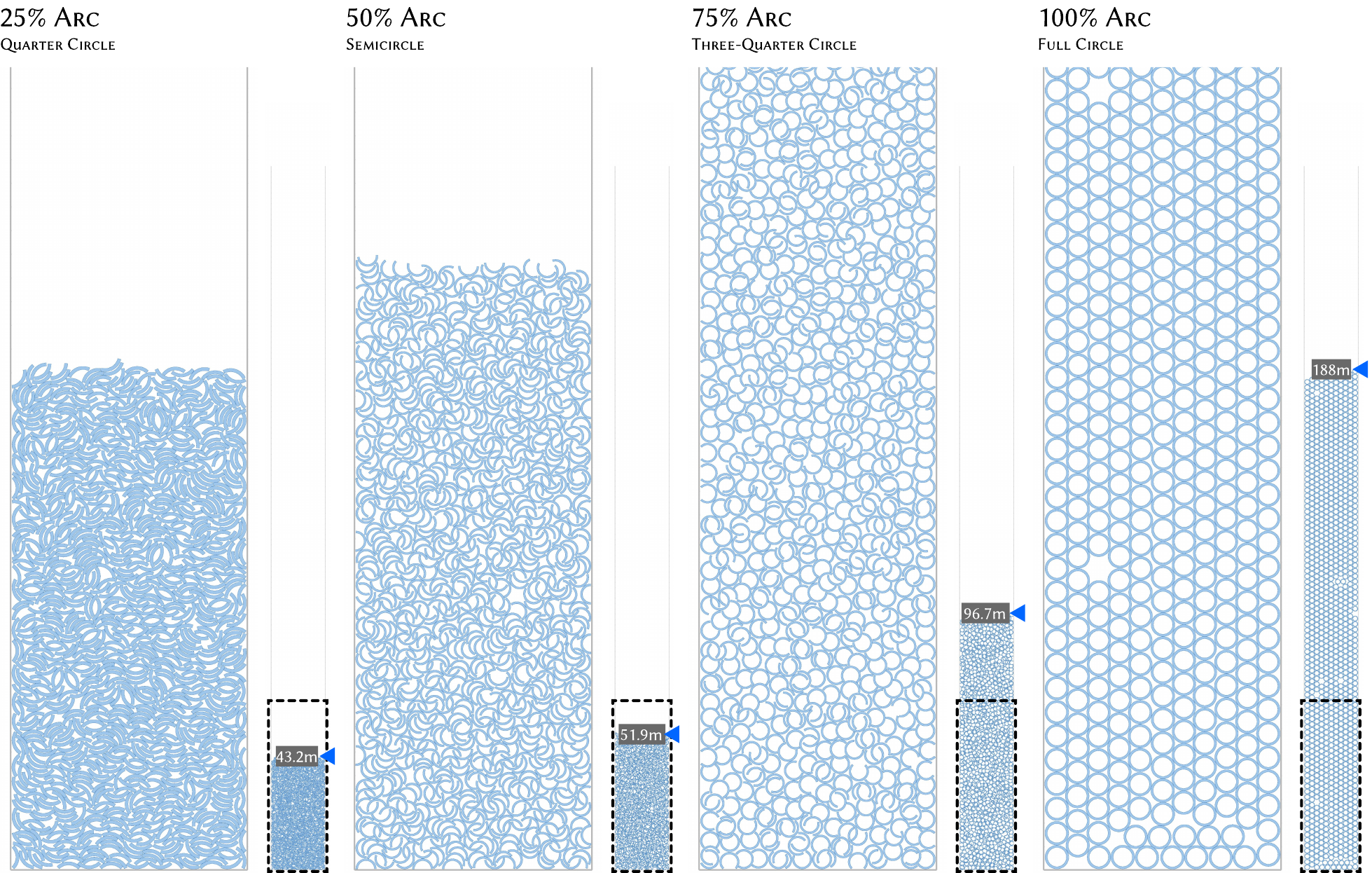}
    \caption{\textbf{Impact of Non-convexity on Packing}. We release $1000$ arc-shaped grains with increasing convexity (sampled from a $0.7$m-radius circle at $25\%$, $50\%$, $75\%$, and $100\%$) into the same vertical column.}
    \label{fig:column_packing}
\end{figure}

\reffig{fig:column_packing} shows that the final packed column height increases dramatically as the grain shape becomes more convex. For the $25\%$ and $50\%$ arcs, grains readily nest and interlock by occupying each other’s interior regions, leading to dense packing and relatively short columns. At $75\%$, this nesting mechanism is largely lost; while grains can occasionally interpenetrate each other’s concave regions in specific configurations, such arrangements are rare, resulting in a significantly lower packing fraction and a much taller column. For the $100\%$ case (full circles), access to the interior region is completely eliminated, preventing any interlocking and yielding the loosest packing and tallest column. These results demonstrate that increasing convexity alone, under identical contact parameters, strongly reduces packing efficiency.

\subsection{Spinning Drum}

To explore the effect of high polydispersity in grain shape, we drop a collection of $9$ grain shapes into a rotating drum. See \reffig{fig:spinning_drum}. To enable this simulation, we trained pairwise maps between each grain type, including the objects used to build the drum.
Despite there being 10 grain types, and a quadratic scaling in the number of networks, we use small $3 \times 32$ networks that take less than 20kB of storage and less than 25 seconds to train, making the simulation quite scalable.

\subsection{Cereal Pouring}
\begin{figure}[ht]
    \centering
    \includegraphics[width=\columnwidth]{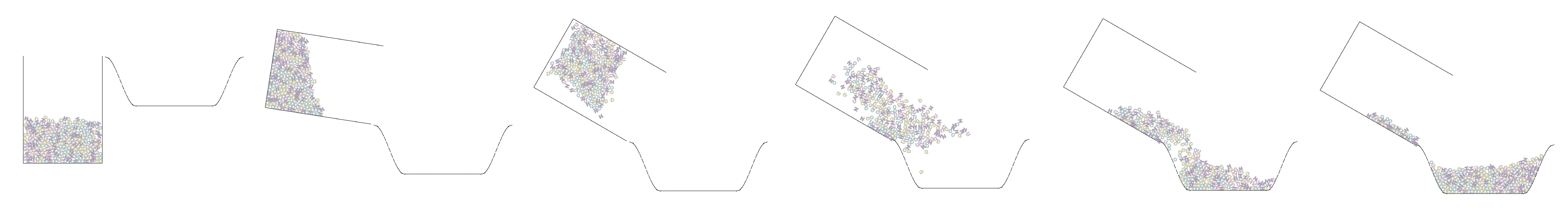}
    \caption{\textbf{Breakfast cereal} poured into a curved bowl. We simulate the cereal with U, D, O, and N shaped grains, mimicking the non-convex shapes found in common breakfast cereals.}
    \label{fig:cereal_pouring}
\end{figure}

Many foods are composed of granular materials, including breakfast cereals. The varied and non-convex shapes of the grains in these cereals make them an interesting real world test case for our method. We prepare a bowl of breakfast cereal, with grain shapes U, D, O, and N, by tipping a box of these grains into a curved container. See \reffig{fig:cereal_pouring}. This test shows that we can build a collection of maps to support another varied family of grain shapes, and that we can support complex boundary shapes by constructing them from a base boundary building block.

\subsection{Coiling Chain}

\begin{figure*}[ht]
    \centering
    \includegraphics[width=\linewidth]{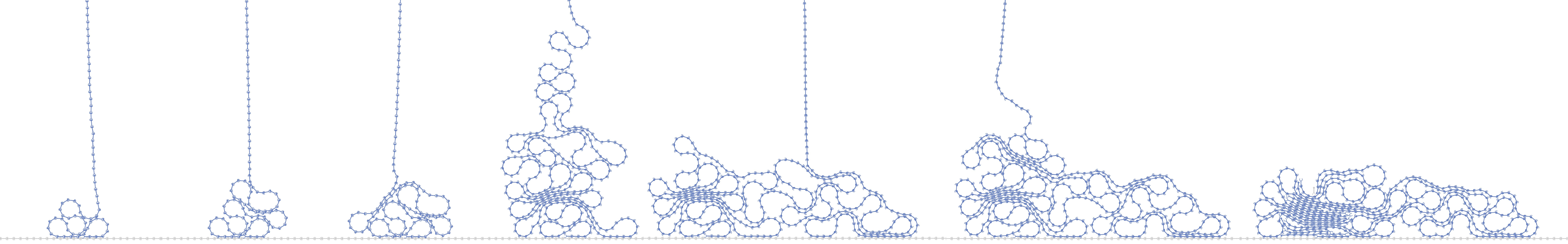}
    \caption{\textbf{Coiling Linkage Rods.} We lower a chain built with 500 rigid links onto the floor. We can easily model a ball and socket joint with non-convex geometry, and our contact formulation only generates a single force per joint. This unified formulation allows us to easily simulate the coiling of the chain as it interacts with the floor and itself.}
    \label{fig:chain_drop}
\end{figure*}

While we primarily target granular media in our validations and experiments, our method applies to a wider family of rigid body simulation applications. As an example, we assemble a chain from individual rigid bodies that connect to neighboring bodies through a ball and socket style geometry, see \reffig{fig:chain_drop}. We do not enforce these joint constraints with an explicit ball and socket joint, but instead with interlocking in the non-convex geometry of the individual chain links. These joints do not impose any additional cost or complexity on the simulation algorithm beyond a single force per joint.

%% file: sections/discussion.tex
\section{Discussion and Conclusion}

We have presented a granular media simulation method that uses learned configuration space distance maps and projected moment arm maps between pairs of grains to accelerate the narrow phase of the simulation and to support heterogeneous, non-convex grain types.
Our results are currently limited to two-dimensions, however, and an extension to three-dimensions would expand its applicability to more real-world problems.
We are also interested in extending the approach from discrete object pairs to \textit{continuous} object pairs, parameterized by a latent code.
Neural networks~\cite{park2019deepsdf} can successfully represent these classes of shapes, and could be extended to encode configuration-space distance fields.
Similarly, we could expand the range of simulations supported by our method by developing specialized grain-boundary maps, where the difference in scale between the boundary and grains leads to significantly different interaction behaviours~\cite{lai2022machine}.
Currently, we model boundaries either as static collections of grains, as flat planes, or as bounding circles, so higher-fidelity boundary models could improve simulation results.

Beyond improvements in geometric capability, we are also interested in improving the quality of the simulation and the network output.
While the penalty-based solver we use is common in the mechanics literature, our neural maps merely provide geometric information and do not impose any particular force model on the solver, and in principle any granular media solver is compatible with our formulation.
LCP-based solvers are known to be less sensitive to parameter tuning, and have been used for granular media~\cite{moreau1994granular}, sometimes in conjunction with an iterative Gauss-Seidel solver~\cite{Smith:2012}.
Rigid body simulations often use continuous collision detection (CCD) to avoid tunnelling artifacts~\cite{ferguson2021intersection,lan2022affine}, which would be a complementary addition to an improved solver along with adaptive timestepping. In the discrete linkage test, for example, if instead of steadily lowering the chain we instead drop the linkage on the floor, the violent impact causes the ball and socket joints to tunnel through one another, snapping the chain.
\begin{wrapfigure}{r}{0.2\textwidth}
\centering
\includegraphics[width=0.2\textwidth]{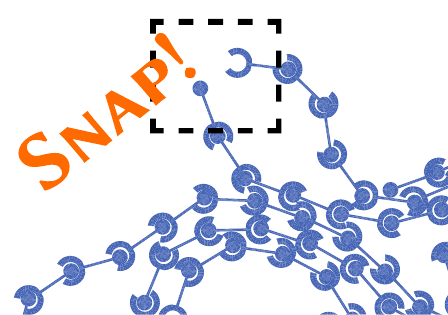}
\end{wrapfigure}
It would also be interesting to explore ways in which we could relax or even lift the single-foot friction formulation, potentially with an improved model that gracefully incorporates several moment arms in a consistent manner with the rotation gradient.
Finally, we are interested in improving our neural distance fields with a better sampling method, particularly near the zero level set in configuration space.
This problem has many approximate solutions for implicit surfaces embedded in Euclidean space, but the setting of higher-dimensional configuration space is far less explored.
Analogous approaches from Euclidean surface reconstruction, particularly in the presence of unstructured or sparse point samples~\cite{sellan2023reach}, could produce explicit hypersurfaces that we could easily sample, taking care to understand and handle new types of gradient singularities that might emerge, such as discontinuous dimensionality reduction (e.g., a screw entering a fitted nut and losing a translational degree of freedom).


%% file: sections/appendix/appendix.tex
\clearpage

\appendix
\section{Contact Force Algorithm}
\label{app:alg}

\begin{algorithm}[!ht]
\caption{$\SEtwo{}$ C-Space Penalty Contact}
\DontPrintSemicolon
\SetAlgoLined
\KwIn{penetration depth $d$; rigid body configurations $q_A,q_B\in\SEtwo{}$; rigid body velocities $v_A,v_B \in \mathbb{R}^3$;
      config gradient $n$; projected moment arms $(\mathrm{proj}\,r_A,\mathrm{proj}\,r_B)$;
      contact force parameters $(k_n,\gamma_n$; friction force parameters $k_t,\gamma_t,\mu)$; spring length $s$; time step size $\Delta t$}
\KwOut{$\mathbf{f}_{n,A},\mathbf{f}_{t,A},\mathbf{f}_{n,B},\mathbf{f}_{t,B}$; updated $s$}

\BlankLine
\textbf{Frames:} $n_{\mathrm{cfg}}\gets n$; $n_{\mathrm{cfg},1:}\gets \mathrm{normalize}(n_{1:})$\;
$R\gets\mathrm{RotMat}(q_A[0])$;\ $n_w\gets R\,n_{\mathrm{cfg},1:}$;\ $t_w\gets[-n_{w,y},\,n_{w,x}]^T$\;

\BlankLine
\textbf{Normal:} $J_A\gets \partial q_{AB}/\partial q_A$;\ $J_B\gets \partial q_{AB}/\partial q_B$\;
$v_n \gets (J_A n_{\mathrm{cfg}})\!\cdot v_A + (J_B n_{\mathrm{cfg}})\!\cdot v_B$\;
$f_n^{\mathrm{tot}}\gets k_n d - \tfrac{1}{2}\gamma_n v_n$\;
$\mathbf{f}_{n,A}\gets J_A(f_n^{\mathrm{tot}} n_{\mathrm{cfg}})$;\quad
$\mathbf{f}_{n,B}\gets J_B(f_n^{\mathrm{tot}} n_{\mathrm{cfg}})$\;

\BlankLine
\textbf{Tangential:} $v_t \gets t_w\!\cdot (v_{B,1:}-v_{A,1:}) + (\mathrm{proj}\,r_B)v_{B,0} + (\mathrm{proj}\,r_A)v_{A,0}$\;
$s \gets s + v_t\Delta t$\;
$f_t^{\mathrm{trial}}\gets -k_t s - \tfrac{1}{2}\gamma_t v_t$\;
$f_{\max}\gets \mu |f_n^{\mathrm{tot}}|$\;
\uIf{$\tfrac{1}{2}\gamma_t |v_t| > f_{\max}$}{
  $s\gets 0$;\ $f_t\gets \operatorname{sign}(-v_t)\,f_{\max}$ \tcp*{slip}
}
\uElseIf{$|f_t^{\mathrm{trial}}| > f_{\max}$}{
  $\lambda \gets \textsc{ProjectFactor}(s,v_t,k_t,\gamma_t,\mu,f_n^{\mathrm{tot}})$ \tcp*{$\lambda\in[0,1]$}
  $s\gets \lambda s$;\ $f_t\gets -k_t s - \tfrac{1}{2}\gamma_t v_t$
}
\Else{$f_t\gets f_t^{\mathrm{trial}}$}
$\tau_A\gets-(\mathrm{proj}\,r_A)f_t$;\ $\tau_B\gets-(\mathrm{proj}\,r_B)f_t$\;
$\mathbf{f}_{t,A}\gets[-\tau_A,\ -f_t t_{w,x},\ -f_t t_{w,y}]^T$;\
$\mathbf{f}_{t,B}\gets[-\tau_B,\ \ \,f_t t_{w,x},\ \ \,f_t t_{w,y}]^T$\;

\Return{$\mathbf{f}_{n,A},\mathbf{f}_{t,A},\mathbf{f}_{n,B},\mathbf{f}_{t,B},s$}\;
\end{algorithm}

%% file: sections/appendix/multi_contact_theory.tex
\section{Multi-Point Contact Analysis}
\label{app:multi_contact}

Here we provide more analysis on the case of multi-point contact, expanding on the single-point contact descriptions in the main paper.
Ultimately, the geometric interpretation of the gradients is similar between both cases.

Just like in the main text, we want gradients of $d_{AB}$ near the zero level set, and we discuss the gradient for body $B$ ($A$ is similar).
Translation gradients, denoted $\dx d_{AB}$, provide a direction for normal forces; and rotation gradients, denoted $\dtheta d_{AB}$, provide a direction and moment arm-scaling for normal torques.
There is no straightforward analytic approach to derive these gradients from \refeq{eq:dist_opt}, so we will examine the gradients geometrically, for configurations near the zero level set of $d_{AB}$.
For expositional simplicity, we will examine a case where $A$ and $B$ are touching but not intersecting (on the zero level set of $d_{AB}$), but the same descriptions apply with only minor changes in terminology (e.g., ``contact point'' to ``closest point'') for configurations with small positive and negative distances.

\subsection{Translation Gradients}
In the case of a single contact point, the translation gradient is simply the (Euclidean) contact normal at the closest point, but the situation is more complex when there are $k > 1$ contact points.
Technically, the multi-contact case is not differentiable, but we can define a geometrically well-defined ``gradient'' as the unit-length direction that maximally increases $d_{AB}$, denoted $\dx d_{AB}$.
Such a direction exists in the intersection of positive halfspaces defined by each contact normal $\mathbf{n}_i$, $\mathcal{C}_i = \{ \mathbf{z} \mid \mathbf{z} \cdot \mathbf{n}_i > 0 \}$, and if $\cap_i \mathcal{C}_i = \emptyset$, $\dx d_{AB} = 0$.
This direction does not generally correspond to an individual contact normal when $k > 1$, but moving along $\dx d_{AB}$ still ensures that all contact points separate from $B$, and so it can be used to abstract away the individual contacts into a single vector.
Also, due to translational symmetry, $\dxA d_{AB} = -\dx d_{AB}$.

Although it seems impractical to compute $\dx d_{AB}$ in multi-contact cases when differentiation is not well-defined, it is worth noting that it coincides with a smoothed distance function which does have well-defined gradients.
The particular choice of smoothing kernel does not matter as long as it is isotropic and spatially decaying or compactly supported, so smoothing does not impact regions that are already smooth and eliminates discontinuities in a geometrically consistent manner.

\subsection{Rotation Gradients}
The single-contact situation is more complex for rotation gradients than for translation gradients.
Rotation gradients depend on the moment arm $\mathbf{r}_1 = \mathbf{p}_1 - \mathbf{x}_A$ from the center of mass of $A$, $\mathbf{x}_A$, to the contact point $\mathbf{p}_1$.
The moment arm can be decomposed into a normal component and a tangential component, based on projection onto the translation gradient $\dx d_{AB}$ which we will denote by $\mathbf{n}$ for simplicity: $\mathbf{r}_1^\perp = (\mathbf{r}_1 \cdot \mathbf{n}) \mathbf{n}$ and $\mathbf{r}_1^\parallel = \mathbf{r}_1 - \mathbf{r}_1^\perp$, or equivalently $\mathbf{r}_1^\parallel = (\mathbf{r}_1 \cdot \mathbf{t})\mathbf{t}$ where $\mathbf{t} = R(\pi/2)\mathbf{n}$ is the tangent direction at $\mathbf{p}_1$.
An infinitesimal counterclockwise rotation induces a change in $\mathbf{p}_1$ and hence $\mathbf{r}_1$ as well.
This infinitesimal change is along a 90-degree counterclockwise rotation of $\mathbf{r}_1$; retaining the decomposition with respect to $\mathbf{n}$, this gives the change as $R(\pi/2)(\mathbf{r}_1^\perp + \mathbf{r}_1^\parallel) = (\mathbf{r}_1 \cdot \mathbf{n})\mathbf{t} - (\mathbf{r}_1 \cdot \mathbf{t})\mathbf{n}$.
Thus, the change in the normal direction is $-\mathbf{r}_1 \cdot \mathbf{t}$ and the change in the tangent direction is $\mathbf{r}_1 \cdot \mathbf{n}$.
The perturbation in the tangential direction does not change the distance, so the component that increases distance must be $-\mathbf{r}_1 \cdot \mathbf{t}$ along $\dx d_{AB}$.
For the multi-contact case, the situation is similar to the translation gradient where $\dtheta d_{AB}$ must be compatible with the single-point rotation gradients.
If different contacts have opposing signs, then $\dtheta d_{AB} = 0$; otherwise, it is the smallest-magnitude value of $-\mathbf{r}_i \cdot \mathbf{t}$ among all contacts with moment arms $\mathbf{r}_i$.